\documentclass[]{aipproc}

\layoutstyle{6x9}

\SetInternalRegister\hbadness{8000} 

\usepackage{epsfig}
\usepackage[scriptsize]{subfigure} 

\newcommand{\reaktion}{\mbox{$pp\rightarrow\,ppK^+K^-$}}

\begin{document}

\title 
      [Associated strangeness production in pp collisions]
      {Associated strangeness production in pp collisions near threshold}

\classification{13.60.Hb,13.60.Le,13.75.-n,25.40.Ve}
\keywords{Document processing, Class file writing, \LaTeXe{}}

\newcommand{\ikpjuel}{IKP, Forschungszentrum J\"{u}lich, D-52425 J\"{u}lich, Germany}
\newcommand{\ikpmue}{IKP, Westf\"{a}lische Wilhelms--Universit\"{a}t, D-48149 M\"{u}nster, Germany}
\newcommand{\cracow}{M.~Smoluchowski Institute of Physics, Jagellonian University, PL-30-059 Cracow, Poland}
\newcommand{\nphycracow}{Institute of Nuclear Physics, PL-31-342 Cracow, Poland}
\newcommand{\catowice}{Institute of Physics, University of Silesia, PL-40-007 Katowice, Poland}
\newcommand{\zeljuel}{ZEL,  Forschungszentrum J\"{u}lich, D-52425 J\"{u}lich,  Germany}

\author{P.~Winter for the COSY-11 collaboration}{address={\ikpjuel}}

\copyrightyear  {2001}

\begin{abstract}
Motivated by the ongoing discussion concerning the nature of the scalar resonances $f_0(980)$ and $a_0(980)$, the COSY-11 collaboration has taken exclusive data on the \reaktion\ reaction near the production threshold. A first total cross section $\sigma=(1.80\pm0.27^{+0.28}_{-0.35})$\,nb for the excess energy $Q=17\,$MeV has been determined. In contrary to the $\eta,\ \omega$, and $\eta'$ single meson production studies which clearly show the strong $pp$ final state interaction (FSI), the cross section values obtained at COSY-11 and DISTO can be both described by a fit with a four-body phase space including the proton-proton final state interaction as well as with one-meson exchange calculations neglecting FSI effects. Therefore, one might think about a compensation of the strong $pp$ interaction through a $pK^-$ FSI effect or an additional degree of freedom caused by the four-body final state. In the latter case, strong FSI  effects can be expected at $Q$-values very close to the $K^+K^-$ production threshold. Such a motivation triggered -- in combination with the investigation of the $K\bar{K}$ interaction being relevant to the structure of the $f_0(980)$ -- further measurements at the excess energies $Q=10$ and $Q=28\,$MeV at COSY-11.\\
\end{abstract}

\date{\today}

\maketitle

\section{Introduction}

Meson production close to threshold in nucleon-nucleon collisions offers an excellent tool in order to study meson-meson and baryon-meson interactions. Due to the low excess energies, the outgoing particles have a low relative momentum and hence final state interactions are more pronounced giving the possibility to derive e.\,g. scattering parameters. Additionally, because of the few relevant partial waves, the theoretical description of the reaction is simplified. Furthermore, studies of these elementary production processes enable to learn about the underlying production mechanisms. Last but not least, the high momentum transfer in such reactions probes the short range components of the hadronic interactions.

Over the last years, there have been several experimental investigations on the close to threshold meson production at different accelerators covering a mass range from the $\pi$- up to the $\phi$ meson. Recently, first results of the challenging studies of the close to threshold cross sections for the production of the broad $f_0(980)$\footnote{In the following, $f_0$ and $a_0$ shall be regarded as an abbreviation for $f_0(980)$ and $a_0(980)$, respectively.} and $a_0^+$ resonances via the proton-proton collisions have been reported \cite{moskal:03,kleber:03}. The COSY-11 collaboration studies -- besides other reactions -- the $K^+K^-$ production in proton-proton collisions. A first cross section $\sigma(Q=17\,\mbox{MeV})=(1.80\pm0.27^{+0.28}_{-0.35})$\,nb has been published~\cite{quentmeier:01-2}. Subsequently, the measurements have been extended to excitation energy values of $Q=10$ and $Q=28$\,MeV.

\section{Experiment}
\begin{figure}[h]
\rotatebox{-90}{\epsfig{file=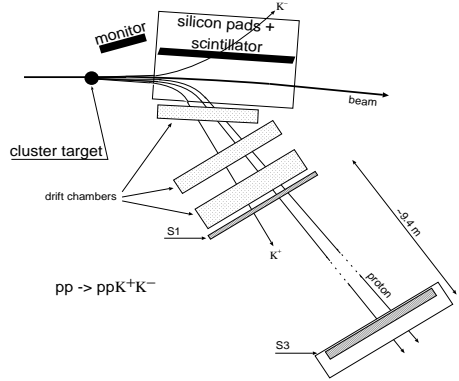,scale=0.3}}
\caption{Experimental setup of the COSY-11 detection system.\label{c11}}
\end{figure}

The internal experiment COSY-11 \cite{brauksiepe:96} at the COoler SYnchrotron COSY \cite{maier:97nim} is shown in Figure \ref{c11}.  A hydrogen cluster target \cite{dombrowski:97} is mounted in front of one of the regular COSY dipoles which acts as a magnetic spectrometer. The positively charged ejectiles are bent to the inner part of the ring and detected by a set of drift chambers followed by a time of flight (TOF) measurement via two scintillator detectors S1 and S3. The momentum determination is performed by tracing back the reconstructed trajectories through the known magnetic field to the interaction point. Together with the velocity calculable from the TOF measurement the four momentum of the positively charged particles are determined. Therefore, an identification of the $ppK^+$-system via the invariant mass of each track enables to identify the $K^-$ by means of the missing mass method. 
Additionally, in the inner part of the dipole gap, a silicon pad detector in combination with a scintillator detector allows for the detection of negatively charged particles which is essential for a nearly 100\% background reduction as will be shown later.

The luminosity is determined by the simultaneous measurement of the elastic $pp$ scattering \cite{moskal:01}. The detectction of the second proton is performed with another silicon pad detector, the so called monitor detector in Figure~\ref{c11} close to the target.

\section{Physics motivation}
\begin{figure}[h]
\epsfig{file=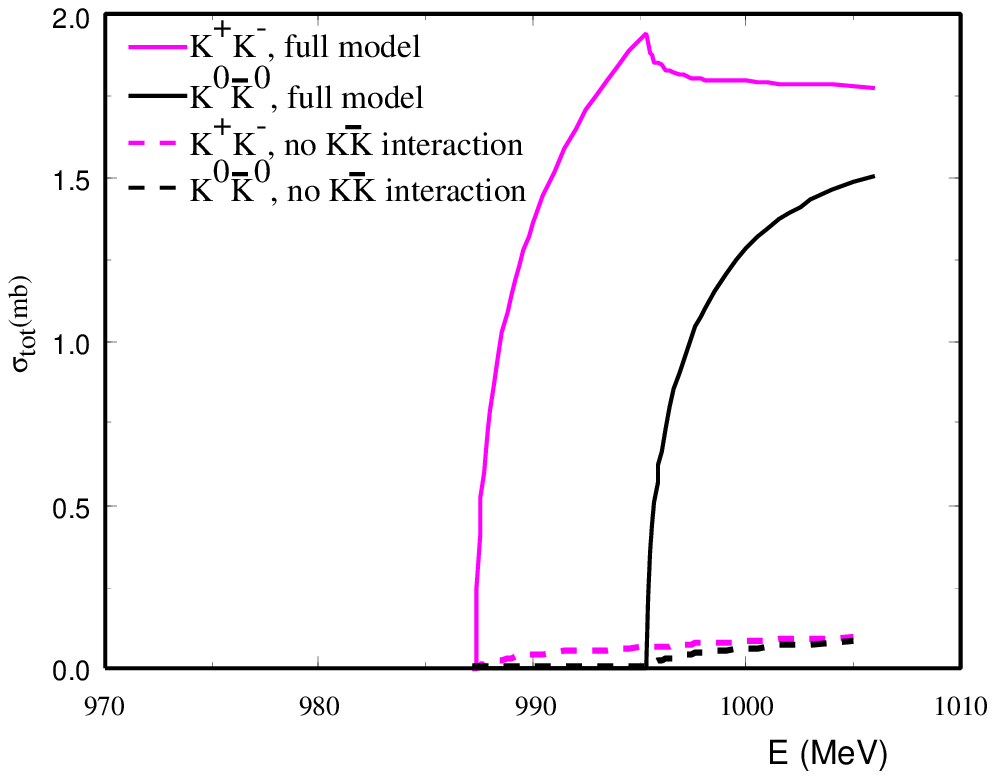,scale=0.65}\hspace{0.8cm}
\epsfig{file=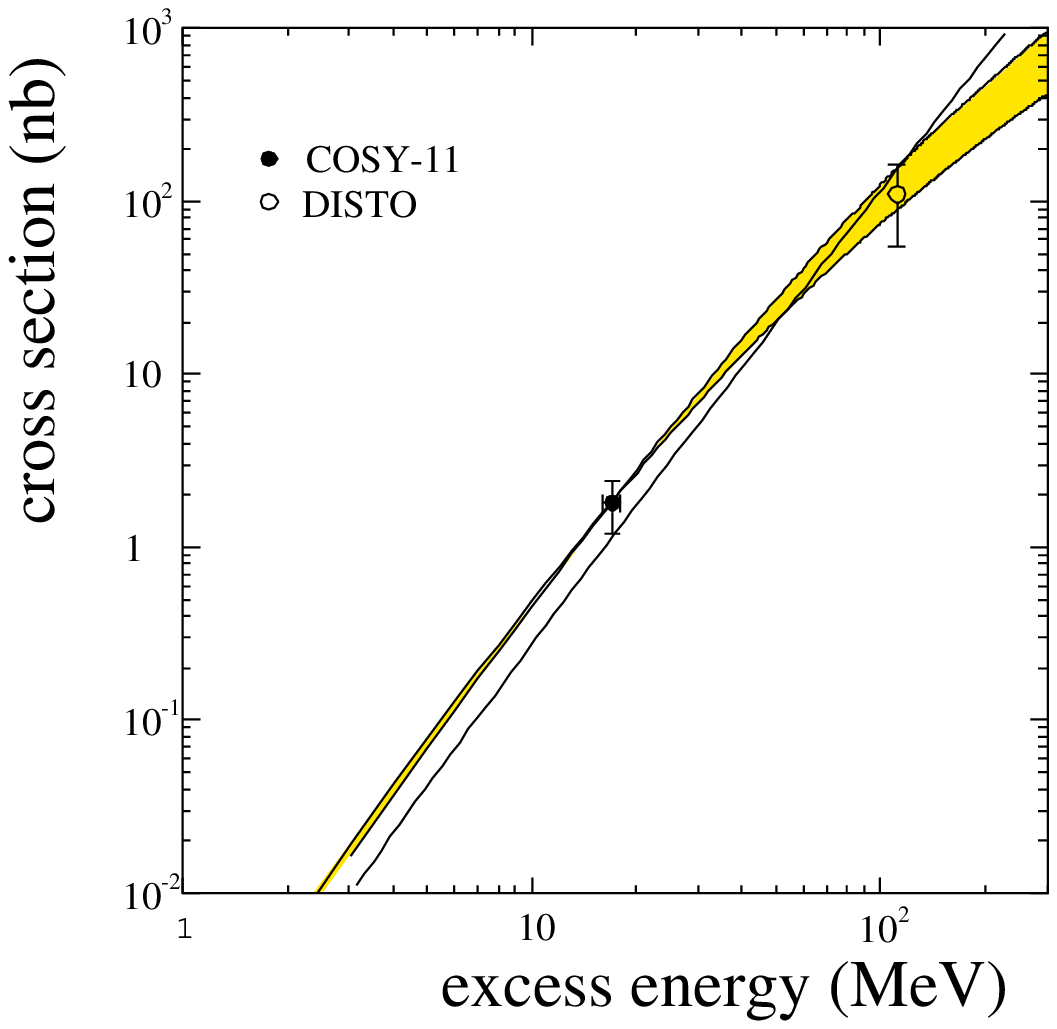,scale=0.5}
\caption{{\it Left picture}: Total cross section calculated within the J\"ulich meson exchange model \cite{krehl:97} for the $\pi^+\pi^-\rightarrow K\bar{K}$ reaction. {\it Right picture}: Total cross section for \reaktion. The data are taken from \cite{balestra:00,quentmeier:01-2} while further information on the lines are given in the text.\label{pipiKK}}
\end{figure}

The $K\bar{K}$ interaction has been under discussion for several years not only in the context of the structure of the isoscalar meson $f_0$. In the framework of a one-boson exchange model several theoretical groups have been working in this field. The strength of the $K\bar{K}$ interaction is crucial for the energy dependence of the cross section $\pi^+\pi^-\rightarrow K\bar{K}$ calculated by Krehl, Rapp and Speth \cite{krehl:97}. Figure \ref{pipiKK} (left part) shows the cross section for both $K^{0}\bar{K}^0$ and $K^+K^-$ production including the $K\bar{K}$ interaction (solid lines) and without (dashed lines). It is obvious that the cross section rises much steeper for the inclusion of the kaon anti-kaon interaction. A similar behaviour can be expected for a proton-proton initial state although complete calculations for this case were not yet performed. 

The knowledge of the $K\bar{K}$ interaction strength will certainly contribute to the understanding of the structure of the $f_0$ meson. Just to give an idea on the still open question concerning the $f_0$ we will mention some models without claiming any completeness.

The interpretation of this resonance as a normal $q\bar{q}$ state within the isoscalar nonet seems to be disfavoured \cite{weinstein:90,morgan:93}. The predictions for the decay width of such states as well as the coupling to pseudoscalar final states (${f_0\rightarrow\pi\pi,}\ {a_0\rightarrow\pi\eta}$) are more than one order of magnitude higher than the experimental observations. The authors of reference \cite{morgan:93} find in their unitarity-enforcing analysis within a Jost function representation of the S-matrix a clear preference for the standard Breit-Wigner description, while in one of their former papers \cite{au:87} the outcome of the analysis was a two resonance description of the $f_0$ meson. The changed result can be attributed to the inclusion of new data on $J/\psi$ and $D_s$ decays. Another prescription has been given by R. J. Jaffe \cite{jaffe:77} where the $f_0$ is assigned as a member of the lightest cryptoexotic $q^2\bar{q}^2$ nonet. The calculations are based on a semiclassical approximation to the MIT bag theory \cite{chodos:74}. Such a $(qq\bar{q}\bar{q})$ configuration rather predicts a rich spectrum of experimentally not confirmed states. This problem does not occur in the potential model by Weinstein and Isgur \cite{weinstein:90} in which the $f_0$ is found to  be a weakly bound $K\bar{K}$ system. Similar findings are given in the framework of the J\"ulich meson exchange model for $\pi\pi$ and $\pi\eta$ interactions \cite{janssen:95,krehl:97}. While the list of such different descriptions could be carried on, it should have become obvious that the structure of the isoscalar mesons is still barely known and that the $K\bar{K}$ interaction plays a crucial role in several models. 

Finally, measurements on the reaction \reaktion\ close to threshold also open the possibility to study final state interactions in the proton-kaon and kaon-kaon systems. Figure \ref{pipiKK} (right side) shows the two existing data points on the total cross section in \reaktion\ from the DISTO- and COSY-11 collaborations \cite{balestra:00,quentmeier:01-2}. The dashed line is a fit of a four body phase space with inclusion of the strong $pp$ FSI. The solid lines are fits of phase space via an intermediate $ppf_0$ state also including the $pp$ FSI. The shaded area stems from the uncertainty of the width of the $f_0$. The present available data do not allow to discriminate between the resonant and nonresonant production. It should be mentioned that also a calculation by Sibirtsev et al. \cite{sibirtsev:97} without any FSI can reproduce the actual cross sections. This led to the speculation that a partial compensation of the $pp$ and $pK^-$ interaction takes place or an additional degree of freedom in the four body final state is responsible for the absence of those FSI effects.

\section{Status of the analysis}
\begin{figure}[h]
\epsfig{file=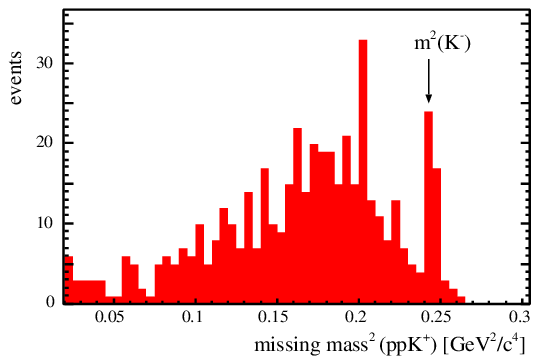,scale=1.1}\hspace{0.6cm}
\epsfig{file=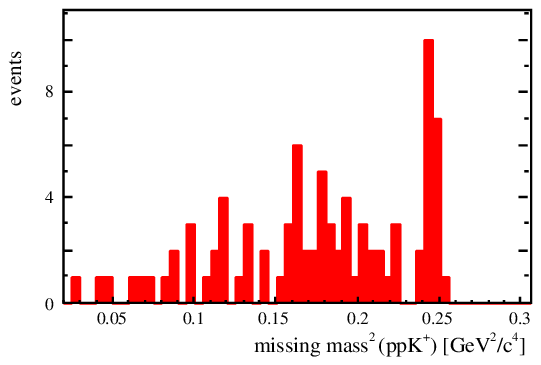,scale=1.1}
\caption{{\it Left picture}: Squared missing mass of the identified $ppK^+$ system. {\it Right picture}: The same events like in the left picture but with an additional hit in the silicon pad detector.\label{missmass}}
\end{figure}

In order to obtain further insight into the understanding of the $K\bar{K}$ interaction, the COSY-11 collaboration has extended their measurements on \reaktion\ to excess energies $Q=10$ and $Q=28$\,MeV. The missing mass spectrum for the excess energy of $Q=28$\,MeV is shown in the left part of figure \ref{missmass}. The selected events include two identified protons and a positive kaon in the final state. 
Besides a clear peak at the corresponding $K^-$ mass a broad background is observed which is understood in terms of the intermediate excitation of the $\Sigma(1385)$ and $\Lambda(1405)$ resonances and misidentified pions stemming from $pp\rightarrow pp\pi^+X$ reactions~\cite{quentmeier:01-2}. It should be emphasized that in the area of the $K^-$ peak, the contamination from the background is very small. The same events with the additional demand for a hit in the silicon pad detector mounted in the dipole gap are shown in the right part of figure \ref{missmass}. While the total amount of \reaktion\ events reduces as expected from Monte Carlo simulations, this cut drastically influences the background structure. Already at this stage the events within the $K^-$ peak are nearly 100\% background free.

\begin{figure}[h]
\epsfig{file=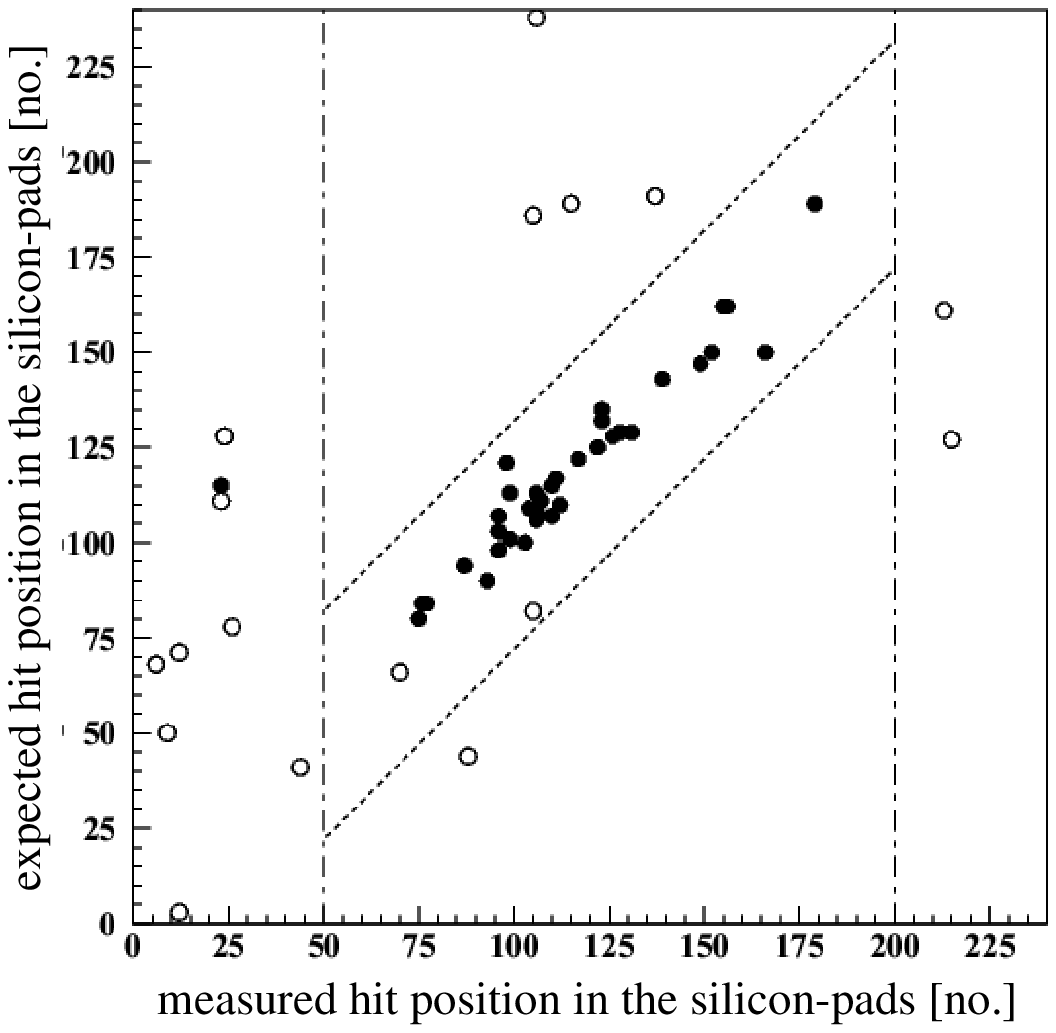,scale=0.37}\hspace{0.6cm}
\epsfig{file=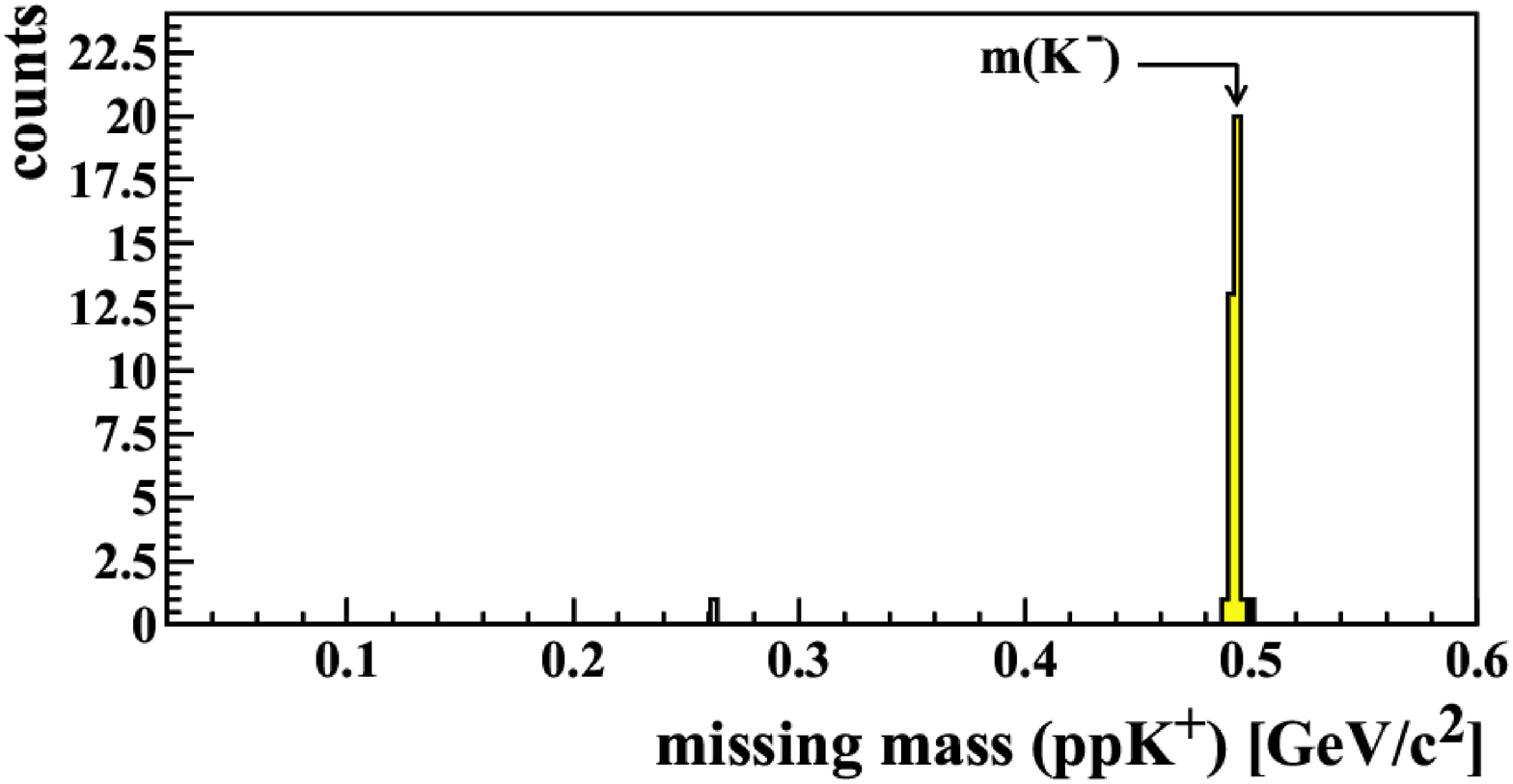,scale=0.23}
\caption{{\it Left picture}: Comparison of the measured hits in the silicon pad detectors for identified $ppK^+$ events. {\it Right picture}: Missing mass spectrum for $Q=17$\,MeV for events with two identified protons, one identified negative kaon and a hit in silicon pads within the cut explained in the text.\label{siliconhit}}
\end{figure}

A last step in this analysis will be not only to demand a hit in the silicon pad detectors but also to compare the hit position with the reconstructed one. Since the four momenta of the 2 protons and the $K^+$ are known, one can calculate the missing four momentum and reconstruct the expected trajectory of the $K^-$ meson in the magnetic field~\cite{quentmeier:01}. Since this procedure was not yet applied to the data set at $Q=28$\,MeV, we will depict its efficiency with the already published data at $Q=17$\,MeV \cite{quentmeier:01-2}. The comparison between the measured and expected hit position in the silicon pads for identified $ppK^+$ events is shown in figure \ref{siliconhit} (left side). The black dots indicate those events which were assigned, according to the experimental resolution, to the $K^-$ peak and the open circles all others. Whereas all events in the kaon signal scatter slighty around the expected correlation, the other events distribute quite randomly over the plot. The applied cuts (dashed lines) were extracted from Monte Carlo simulations in order to simultaneously minimize the loss of $K^-$ events and to maximize the reduction of the background. 
The result of applying this last cut to the selected events is a essential background free missing mass spectrum as is shown in the right part of figure \ref{siliconhit} for $Q=$17\,MeV. It is expected that the same supression of the background will be achieved for the two data sets at $Q=10$ and $28$\,MeV. 

\section{Summary}
The COSY-11 collaboration extended their measurements on the elementary production of $K^+K^-$ in $pp$ collisions with new data at $Q=10$ and $Q=28$\,MeV. From a former measurement at $Q=17$\,MeV it is known, that this can be achieved background free. The analysis of both data sets is still in progress but for the higher excess energy a clear kaon signal is already seen. The new information on the excitation function in this reaction will help to understand the underlying process and the strength of the $K\bar{K}$ interaction. This might help to clarify the open question on the structure of the isoscalar meson $f_0$.


\bibliographystyle{aipproc}
\bibliography{abbrev,general}

\end{document}